# Conversations with Data: How Data Journalism Affects Online Comments in the *New York Times*

Avner Kantor
University of Haifa
avnerkantor@gmail.com

Sheizaf Rafaeli
Shenkar College
sheizaf@rafaeli.net

## Abstract

*Users in the data age have access to more data than ever before, but little is known how they interact with it. Using transparency and multimedia, data journalism (DJ) lets users explore and interpret data on their own. This study examines how DJ affects online comments as a case study of user interactions with data. The corpus comprises 6,400 stories and their comment sections from the DJ and other sections of the* New York Times*, from 2014-2022. Results indicate that DJ is positively associated with higher level of interactivity between the users. This relationship is mediated by statistical information, information sources, and static visualizations. However, there is a low level of interactivity with the content; consequently, only part of the users use it. The results demonstrate how data accessibility through DJ engages the users in conversation. According to deliberation theory, this creates a conducive environment for democratic processes.*

## 1. Introduction

We are living in an era of data revolution, characterized by unprecedented access to vast amounts of data and advanced tools to analyze, interpret, and present it (Aaltonen & Penttinen, 2021). This increased access theoretically allows for more informed decision-making across many aspects of our lives (Lewis & Westlund, 2015). However, in practice, fewer people participate in meaningful conversations, and the gains in access to data seem to be counterbalanced by a decline in argumentative complexity and deliberation (Dryzek et al., 2019). This could be attributed to the difficulty of using data effectively, as it requires cognitive abilities and effort (Lesage & Hackett, 2013). To support and encourage broader and more effective use, it is necessary to understand how people interact with data (Boichak et al., 2019).

News websites are one of the primary sources of public information in the 21$^{st}$ century (Esau et al., 2017). Their comment sections enable the public to respond immediately to information, engage in conversation, and try to affect public deliberation. (Esau et al., 2017; Wang & Diakopoulos, 2021). In spite of their potential, several newspapers have removed comment sections because they had suffered from incivility and lack of rationality (Esau et al., 2017).

In the past decade, data journalism (DJ) has become a major way of facilitating access to data and transparency (Erkmen, 2024). It has gained significant traction in newsrooms across the globe. DJ utilizes various methods to present data, support and engage the public. However, the current understanding of how the public interacts with DJ remains limited, highlighting the need for further quantitative research in this area (Erkmen, 2024).

This study examines how DJ affects user comments as a case study of user interactions with data. Specifically, we examine user comments of DJ and other sections in the *New York Times* (NYT) from 2014-2022. The corpus comprises 6,400 stories and 785,883 comments. Analysis of the comment sections reveals that DJ is positively associated with user engagement in conversation. This relationship is positively mediated by statistical information, information sources, and static visualizations. However, the results show that only part of the users interact with the DJ content, probably those who are with the required cognitive skills and willingness to exert the necessary effort. The study demonstrates the connection between data accessibility and interactivity. It also illustrates the role that news factors and transparency play in the relationship. Based on deliberation theory, this creates an environment in which democratic processes can flourish.

## 2. Literature review

### 2.1. The role of data

Democratic deliberation requires information about social issues (Friess & Eilders, 2015). Thus, information facilitates the development of a collective understanding of problems and methods of addressing them. Citizens who are knowledgeable about opposing arguments may be able to change their opinions based on this information (Coleman & Moss, 2012).

Data in the form of statistics, facts, and figures can provide audiences with insight and knowledge (Aaltonen & Penttinen, 2021; Lesage & Hackett, 2013). However, using it requires data literacy skills, and an interest in the subject matter (Boichak et al., 2019). Because of this, although it can be used to empower the public, data may only be useful to some users (Gurstein, 2011). Several methods have been developed to facilitate the public's engagement with data, as discussed in the following section.

### 2.2. Interactivity with data

Interactivity is elusive concept that can emerge out of various elements (Ziegele et al., 2020). First, it can be facilitated by system design and elements, such as systems with hyperlinks or a comment section. Second, users who grasp the interactive potential can interact with the system and further facilitate interactivity by clicking on hyperlinks, visualizations or posting comments. Third, interactivity can be facilitated when users interact with others based on system design, leave comments, and respond to one another to create conversations.

Among the systems that can facilitate interactivity is the comment section in news websites (Weber, 2014). The section enables the audience to post online in response to the content of news reports or to comments thereto made by other users (Wang & Diakopoulos, 2021). This can lead to the creation of public opinion and enable citizens to make informed decisions (Ziegele et al., 2020). However, it was found that commenters are driven more by social motivations than by deliberative ones, which results in only partial cognitive gratification (Springer et al., 2015). Typically, a small and non-diverse group of users contributes most of the comments (Frischlich et al., 2019). This often leads to incivility, low rationality, and echo chambers (Esau et al., 2017). As a result, many news websites have disabled their comment sections (Frischlich et al., 2019).

News factors theory can explain audience interactivity in comment sections (Ziegele et al., 2020). News factors might be defined as relevance indicators applied by both journalists and the audience to determine the newsworthiness of a story (Eilders, 2006). These factors typically include elements like prominence, timeliness, conflict, and facticity, which help gauge a story's potential impact and relevance. Secondary news factors, on the other hand, support the core newsworthiness of the story rather than drive it (Ziegele et al., 2020). In this study, we focus on factors specifically related to facticity, such as statistical information, information sources, and multimedia elements like data visualizations. These aspects are considered secondary news factors because they contribute significantly to the depth, credibility, and analytical value of a story, thereby influencing how users assign importance and engage with the content beyond their personal interests (Ziegele et al., 2020).

**2.2.1 Statistical information** is a traditional method of presenting data to the audience, combined in the news story's content (McConway, 2016). Studies show that it enhances audience perceptions of credibility and quality (Henke et al., 2020; Link et al., 2021).

**2.2.2. Information sources**. Through external hyperlinks, organizational reports, academic articles, and numerical data are presented to the audience. They ensure credibility, quality, and transparency (Lesage & Hackett, 2013). They can also improve credibility assessments and intentions to engage with news (Curry & Stroud, 2021; Link et al., 2021). However, limited effects were observed (Henke et al., 2023; Ksiazek, 2018; Ksiazek et al., 2016; Liu et al., 2015). Conversely, it was found that information sources may negatively impact trust outcomes, as audiences find news stories without them more credible (Tandoc & Thomas, 2017).

**2.2.3. Data visualization**. Multimedia were found to enhance interactivity (Carpenter, 2010; Liu et al., 2015). Specially, data visualization can help in data exploration, comprehension, and interpretation (Link et al., 2021). Finally, visualization was found to be associated with higher levels of news elaboration among high-involved audiences (Lee & Kim, 2016).

### 2.3. Data journalism

In the last decade, leading newspapers have provided sections with massive amounts of data (Erkmen, 2024). DJ offers the audience access to data in various ways such as statistical information, information sources, and data visualization (Stalph, 2018). This transparency allows the audience to better understand the basis of news stories and fosters trust in the reporting process (Zamith, 2019). Additionally, DJ gives the audience an opportunity to interact with data and journalists, to collaborate with other citizens, and to rebuild trust in news media (Boyles & Meyer, 2016; Felle, 2015). Audience interactivity can occur before, during, and after the production of stories by suggesting ideas, collecting data, and analyzing the results (Palomo et al., 2019).

The literature on audience engagement with DJ has grown significantly in recent years (Tong, 2024). From the journalists' perspective, early interviews with data journalists have found that they are passionate about engaging audiences and having data-driven conversations with them (Felle, 2015). However, in recent years, there has been a shift away from the optimistic view of DJ as a means of promoting audience interactivity with data and attracting new audiences (Martin et al., 2024). Data journalists have stated that the audience shows low levels of interactivity with their stories.

From the audience's perspective, interviews indicate an appreciation for the way DJ provides analysis, statistical information, narrative structure, and visualizations (Stalph et al., 2023). An analysis of user comments on the *Economist*'s Graphic Detail blog revealed that the conversation topics included the story's visualization, content framing, and relations with external data (Hullman et al., 2015). Similarly, an analysis of comments on climate change stories in the *Breitbart News*, *Guardian*, and NYT found that the audience discussed the stories' content, but this occurred in only a small percentage of the comments (McInnis et al., 2020). Furthermore, a study on the *Guardian* found that the audience is more engaged with DJ than with traditional journalism, as indicated by the higher ratio of replies to comments (Kantor & Rafaeli, 2021). However, in contrast to these findings, several studies indicate that audiences do not interact with data visualizations, the data, or the annotation layer (Boy et al., 2015; Felle, 2015). These mixed findings highlight the need for further quantitative research in this area.

## 3. Research model

This study examines DJ and user comments in order to extend the understanding of data's role in deliberation. It was found that commenters are driven more by social motivations and less by cognitive gratification (Springer et al., 2015). This finding might contradicts data journalists' hopes for fostering data-driven deliberation (Felle, 2015). To provide further evidence of DJ's contribution to deliberative processes, we developed a conceptual model.

Figure 1 presents a model that includes the main characteristics of DJ and audience interactivity within the comment section. The model structure allows for comprehensive examination of the relationship between all elements. The empirical results will provide a deeper understanding of the underlying mechanisms and their effects on deliberation. The hypotheses underlying the model are described below.

Several studies have hypothesized that DJ can foster conversations (Felle, 2015; Kantor & Rafaeli, 2021).

**H1**. *DJ is positively related to conversation (interactivity between users).*

DJ is characterized by the extensive use of elements designed to attract, support, and engage the audience (Stalph, 2018). Audiences were found to be interested in using these elements (Stalph et al., 2023). Such elements function as secondary news factors and are expected to play a positive mediating role (Ksiazek, 2018).

**H2**. *Secondary news factors mediate the relationship between DJ and conversation. DJ is positively related to statistical data (H2a), which is in turn positively related to conversation (H2b). DJ is positively related to information sources (H2c), which is in turn positively related to conversation (H2d). DJ is positively related to static visualizations (H2e), which in turn is positively related to conversation (H2f).*

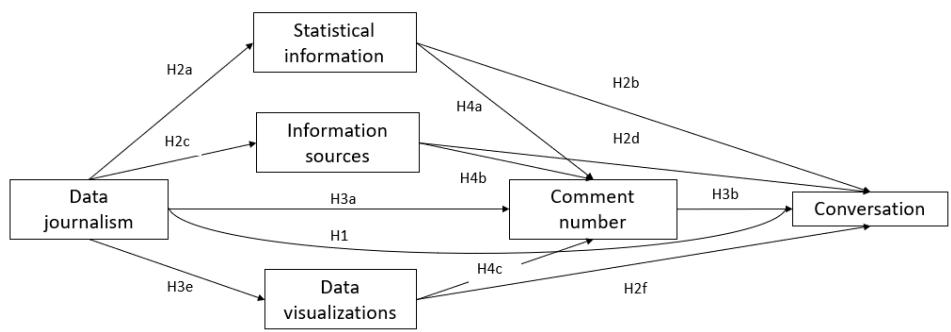

**Fig. 1. Proposed mediation model**

Comment number is often associated with user interest (Ksiazek, 2018). DJ is presumed to attract an audience, and, as such, is expected to increase the comment number. In turn, a higher comment number is likely to foster more conversations (Ksiazek et al., 2015).

> **H3**. *Comment number (interactivity between comment section and user) mediates the relationship between DJ and conversation. Specifically, DJ is positively correlated with comment number (H3a), which in turn is positively correlated with conversation (H3b).*

Assuming Hypothesis 3 is correct, DJ elements are likely to contribute to the comment number. Accordingly, DJ elements and comment number are serially mediate the relationship between DJ and conversation.

> **H4**. *Paths from secondary news factors to comment number mediate the relationship between DJ and conversation. Specifically, there is a positive path from statistical information to comment number (H4a), a positive path from information sources to comment number (H4b), and a positive path from static visualizations to comment number (H4c).*

## 4. Method

### 4.1. Data source

The study data were collected from the NYT, a primary source of public information and one of the oldest and most prestigious newspapers in the world (Wang & Diakopoulos, 2021). NYT stories are widely read by the US elite and have the power to influence the national agenda (Kiousis, 2004). In the current era of fake news, the NYT defends the legitimacy of journalism and its role as the fourth estate (Lischka, 2019). Furthermore, it has a long history of incorporating data visualizations in its stories. In August 2010, a dedicated DJ section was included on the NYT website, starting with Nate Silver's FiveThirtyEight blog. In March 2014, it was replaced by The Upshot.

The NYT allowed commenting on many of its stories since 2007.[1] Before they are published, the comments are manually moderated. According to the NYT, comments that are "articulate, well-informed remarks that are relevant to the article" are accepted.[2] This objective is supported by a study finding that most participants focus on the topic of the story and engage in respectful conversations to argue their points (Ruiz et al., 2011). Furthermore, since comment threads are typically open only for 24 hours,[3] a delay effect on the conversation is avoided. These factors make the NYT a suitable source for studying user interactivity in comment sections.

### 4.2. Data collection

Multiple methods were utilized to collect the study data. First, the NYT archive API[4] was employed to gather all the Upshot stories published between April 2014 and April 2022 ($N = 4,913$). Web crawling was then used to collect the full text and elements of these stories. To collect the stories' comments, the NYT community API and a web scraper were employed.[5] Stories with less than ten comments were excluded to avoid bias. The resulting Upshot sample comprised 3,224 stories.

Second, we determined topic labels for Upshot stories that were missing them using a supervised learning technique. First, we used the official archive API to gather a random sample of NYT stories from different categories such as politics, world, business, sports, culture, education, health, and science. The story label was determined based on the NYT's news desk, section, or subsection. Next, we trained a multinomial naive Bayes model on a labeled sample of 13,500 story abstracts. The model achieved a .85 accuracy on the train-test data, with a precision and recall of .85. Using this classifier, we labeled the entire sample of stories. The resulting topic distribution of the stories included 54% business, 22% politics, and 11% health stories.

Third, stories from other NYT sections were collected. Since commenting is not enabled on most stories outside The Upshot, the number of stories in some topics was not equal. Overall, we collected 6,400 stories and 785,883 comments. Topics with a small number of stories were labeled "other topics". Table 1 presents the final distribution.

---

[1] Hoyt, C. (2007) 'Civil discourse, meet the internet', *New York Times*, 4 November. Available at: https://www.nytimes.com/2007/11/04/opinion/04pubed.html (Accessed: 12 June 2024)

[2] *Comments* (no date) *The New York Times*. Available at: https://help.nytimes.com/hc/en-us/articles/115014792387-Comments (Accessed: 12 June 2024)

[3] Etim, B. (2017) 'The Times sharply increases articles open for comments, using Google's technology', *New York Times*, 13 June. Available at: https://www.nytimes.com/2017/06/13/insider/have-a-comment-leave-a-comment.html (Accessed: 12 June 2024)

[4] *Developer Network* (no date) *The New York Times*. Available at: https://developer.nytimes.com (Accessed: 12 June 2024)

[5] Pietz, T. (2021) 'nytimes-scraper'. Available at: https://pypi.org/project/nytimes-scraper (Accessed: 12 June 2024)

Table 1. Story distribution by type and topic

|  | DJ (%) | Other (%) | Total (%) |
|---|---|---|---|
| Business | 1,741 (53.5) | 774 (23) | 2,515 (37.9) |
| Politics | 713 (21.9) | 861(25.5) | 1,574(23.7) |
| Health | 359 (11.0) | 212 (6.3) | 571 (8.6) |
| Opinion | 162 (5.0) | 701 (20.8) | 863 (13) |
| Sports | 107 (3.3) | 356 (10.6) | 463 (7) |
| World | 68 (2.1) | 288 (8.5) | 356 (5.4) |
| Other | 107 (3.3) | 179 (5.3) | 286 (4.3) |
| *Total* | *3,087(100.0)* | *3,313(100.0)* | *6,400(100.0)* |

Fourth, story components were detected through crawling HTML and Cascading Style Sheets (CSS) features, including external hyperlinks and data visualizations. There are 137 stories with interactive visualizations that were excluded because their comment section button is located in a different area of the webpage, and this might bias the analysis. These stories represent two percent of the corpus.

### 4.3. Measurement

The variables were calculated using various methods. Table 2 presents the descriptive statistics.

DJ stories were categorized according to the Upshot label provided by the NYT, which distinguishes between regular and data-driven journalism. During data collection, we observed a number of stories that resembled the Upshot stories, but were not classified as such by the NYT. To maintain consistency, the study followed the original classification. The corpus consists of 3,087 (48.2%) DJ stories.

*Statistical information* was indicated by the frequency of numbers within the story content. The higher the value per story, the greater the likelihood that statistical information was used in the story. Numbers are not necessarily used to represent statistics, but can also be used for dates or names. However, we found a positive correlation between this measure and DJ, supporting its use. A linguistic inquiry and word count (LIWC) dictionary and software were used to detect frequency (Pennebaker et al., 2015).

*Information sources* were calculated based on the number of external hyperlinks in a story. There was no distinction between different types of sources. The corpus consists of 4,891 (74.7%) stories containing 1-64 information sources.

*Static visualizations* were based on web crawling and analysis. We determined whether a story contained static visualizations and their number. The corpus consists of 1,443 (22.5%) stories containing 1-14 static visualizations.

*Comment number* represents the interactivity level of users with the story content. It was measured by the number of top comments for each story (Ksiazek et al., 2015, 2016).

*Conversation* represents the interactivity level between the users. It was calculated by the ratio between responses and comments for each story (Ksiazek et al., 2015, 2016). The number of responses to each story is normalized using this method.

*Control variables* included variables known to affect comment number and conversation: story topic, year of publication, and story length (in words) (Weber, 2014).

Table 2. Descriptive statistics

|  | Min | Max | *M* | *SD* |
|---|---|---|---|---|
| Statistical information | 0 | 20.04 | 2.39 | 1.33 |
| Information sources | 0 | 64 | 3.75 | 4.79 |
| Static visualizations | 0 | 14 | .36 | .86 |
| Comment number (log) | 2.30 | 7.97 | 4.18 | 1.09 |
| Conversation | .00 | 3.71 | .80 | .44 |

### 4.4. Statistical analysis

The unit of analysis is the story ($N = 6,400$), with all dependent variables summarized at story level. To prevent bias, comments with fewer than 25 words were excluded, and replies were also excluded to focus solely on direct reactions to the story's elements. The comment number was log-transformed to normalize the skewed distribution. Each story topic was dummy-coded into separate independent variables.

The hypotheses were tested using Hayes' PROCESS macro model 80 (Hayes, 2022). The model consisted of three parallel paths connecting DJ to conversation, each having two serial mediators. The model was applied to 5,000 bias-corrected bootstrap samples with 95% confidence intervals. Significant difference ($p < .05$) occurred when both the lower and upper bounds of the confidence interval did not include zero. The use of PROCESS allows for a detailed examination of mediation effects, providing a clear understanding of how different variables interact within the model.

## 5. Results

### 5.1. Preliminary analyses

Table 3 shows the Pearson correlations of the study variables. DJ was positively correlated with statistical information, information sources, and static visualizations but negatively correlated with comment number. Most importantly, DJ was positively correlated with conversation, confirming H1.

Table 3. Pearson correlation coefficients of the study variables

|  | 1 | 2 | 3 | 4 | 5 |
|---|---|---|---|---|---|
| 1. Data journalism | 1.00 | | | | |
| 2. Statistical information | .19*** | 1.00 | | | |
| 3. Information sources | .28*** | .02 | 1.00 | | |
| 4. Static visualizations | .25*** | .21*** | .02 | 1.00 | |
| 5. Comment number (log) | -.20*** | -.19*** | .07 | -.05*** | 1.00 |
| 6. Conversation | .13*** | .07*** | .08*** | .10*** | .16*** |

*$p < .05$; **$p < .01$; ***$p < .001$

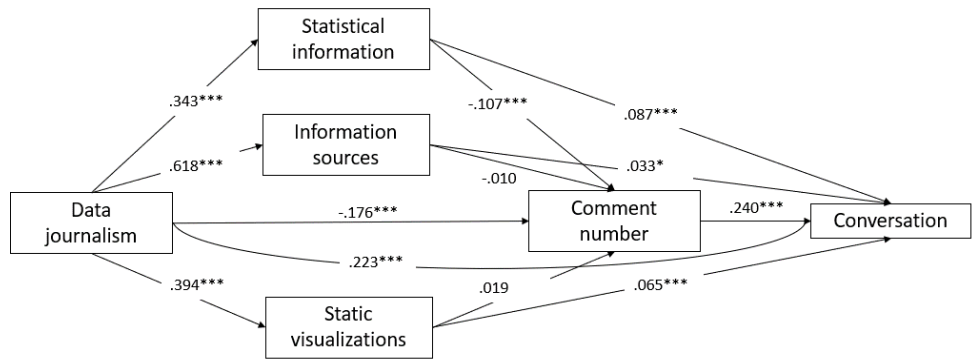

**Fig. 2. Results of multiple serial mediation analysis**

*$p < .05$; **$p < .01$; ***$p < .001$; all effects are standardized

Table 4. Total, direct and indirect model effects

| Model pathways | Effect | SE | LLCI | ULCI |
|---|---|---|---|---|
| Total effect | .111 | .013 | .085 | .136 |
| Direct effect | .099 | .014 | .072 | .125 |
| Partially standardized indirect effects | | | | |
| Total | .027 | .013 | .002 | .053 |
| DJ - Statistical information - Conversation | .030 | .005 | .020 | .040 |
| DJ - Information sources - Conversation | .020 | .008 | .005 | .036 |
| DJ - Static visualizations - Conversation | .026 | .006 | .015 | .037 |
| DJ - Comment number - Conversation | -.042 | .007 | -.057 | -.028 |
| DJ - Statistical information - Comment number - Conversation | -.009 | .001 | -.011 | -.006 |
| DJ - Information sources - Comment number - Conversation | .002 | .002 | -.002 | .005 |
| DJ - Static visualizations - Comment number - Conversation | .002 | .001 | -.001 | .004 |

## 5.2. Mediating effects

Figure 2 shows the results of the serial mediation model. DJ has a significant positive relation with statistical information (supporting H2a) (β=.343, $p<.001$), information sources (supporting H2c) (β=.618, $p<.001$) and static visualizations (supporting H2e) (β=.394, $p<.001$), which in turn has significant paths to conversation (supporting H2b, β=.087, $p<.001$; H2d, β=.033, $p<.05$; and H2f, β=.065, $p<.001$). DJ has a significant negative path to comment number (rejecting H3a) (β=-.176, $p<.001$), which in turn has a significant positive path to conversation (supporting H3b) (β=.240, $p<.001$). Finally, statistical information has a significant negative path to comment number (rejecting H4a) (β=-.107, $p<.001$). The other paths are not significant (rejecting H4b and H4c).

As presented in Table 4, the direct effect of DJ on conversation is significant (β =.111, 95% CI =.085 to .136). The mediation effect of statistical information, information sources, and static visualizations on conversation is also significant (respectively, β = .030, 95% CI = .020 to .040; β =.020, 95% CI =.005 to .036; β =.026, 95% CI =.015 to .037). The mediation effect of comment number on conversation is significant (β =-.042, 95% CI =-.057 to -.028). Finally, the indirect effect of statistical information and comment number as serial mediators in the relation between DJ and conversation is significant (β =-.009, 95% CI =-.011 to -.006). The other paths are insignificant.

## 6. Discussion

The study explores the relationship between data access and audience interactivity. It uses DJ as a case study since it incorporates a variety of methods for presenting data. Audience interactivity is examined through online news comments. The corpus includes 6,400 stories from the NYT and their comment sections. The results indicate that DJ is positively associated with higher level of interactivity between the audience, as it represented by the degree of conversation. A multiple serial mediation model illustrates how statistical information, information sources, and static visualizations contribute to the relationship. However, the interactivity of the audience with the content is low, as evidenced by the comment number. Using the results, the following section examines the role of DJ in relation to theories of transparency, news factors, data visualizations, and deliberation.

## 6.1. Key findings

**6.1.1. DJ and conversation (H1).** The results confirm that DJ is positively associated with conversation, supporting Hypothesis 1. This finding, which refers to interactivity among the audience, aligns with previous research (Hullman et al., 2015; Kantor & Rafaeli, 2021; McInnis et al., 2020) and hopes (Felle, 2015; Martin et al., 2024). The positive association appears consistent across various topics, suggesting that the engaging nature of DJ transcends specific subject matters.

**6.1.2. The mediating role of secondary news factors (H2).** Hypothesis 2 is also supported, as the relationship between DJ and conversation is positively mediated by statistical information, information sources, and static visualizations. This is consistent with previous studies that highlight the audience's interest in these elements (Stalph et al., 2023). The elements require cognitive skills and effort that may be challenging; however, the results indicate that some users do have these capabilities. Although some elements are considered more appealing (Kennedy et al., 2016), no differences are evident in their effect.

**6.1.3. The mediating role of comment number (H3).** Contrary to Hypothesis 3, comment number negatively mediates the relationship between DJ and conversation. This lower interactivity with the content is consistent with the higher interactivity among the audience observed in the H1 results (Ksiazek et al., 2016). It is in accordance with studies of data visualization (Lee & Kim, 2016). In contrast, interviews with news consumers indicate that they are interested in interacting with data (Stalph et al., 2023). This may be explained by DJ's cognitively challenging characteristics. According to gratification theory, it could result in a lower perceived level of interactivity and, therefore, a smaller audience and fewer comments (Ksiazek et al., 2016).

**6.1.4. Secondary news factors and comment number as serial mediators (H4).** The results for Hypothesis 4 indicate that the relationship between DJ and conversation is serially mediated by statistical information and comment number. The negative mediation suggests a lack of digital literacy or personal interest among the general public (Felle, 2015). It is likely that a small group of data experts predominantly consume and interpret the statistical information. The significant total indirect effect suggests that other mediators such as the narrative and emotional impact of the story also affect the relationship (Stalph, 2018).

## 6.2. Interpretation and implications

**6.2.1 Transparency.** It was hypothesized that DJ would have an impact on interactivity based on the connection between data and transparency (Curry & Stroud, 2021; Zamith, 2019). However, previous studies have shown limited effects of transparency (Henke et al., 2023). Based on H2 results we can connect transparency with higher interactivity between the audience and a lower interactivity with the content. Results may indicate that only a portion of the audience is encouraged by transparency. This study contributes to the understanding of transparency's role and can encourage the effort involved in promoting it while providing rare evidence regarding its impact.

**6.2.2. News factors.** The results show that statistical information, information sources and static visualizations play a positive mediating role in facilitating conversation. These findings illustrate their role in engaging the audience and reinforce their designation as secondary news factors. However, statistical information is negatively related to comment number, and the relationship between DJ and conversation is negatively mediated by statistical information and comment number. This suggests that while it can stimulate conversation, it may not encourage the audience to comment. This finding should be taken into consideration when designing news and comments platforms (Esau et al., 2017; Wang & Diakopoulos, 2021).

**6.2.3. Target audiences.** The study results fill a gap that has been identified in the literature regarding DJ impact on target audiences (Erkmen, 2024). The study provides evidence that DJ audiences are associated with a high degree of conversation. Due to the low level of interactivity with the content, the audience is likely to be data experts. Possibly, this is a result of a lack of data literacy or a mismatch between audience interest and willingness to participate. This finding confirms that DJ can enhance the process of public deliberation. It may also suggest that the DJ was not able to reach a wide audience (Lesage & Hackett, 2013).

**6.2.4. Data visualization.** Data visualization was expected to capture users' attention and encourage them to comment (Hullman et al., 2015; Kennedy et al., 2016; Link et al., 2021). However, the results indicate no significant effect on comment number. This can be attributed to several reasons. One possible is lack of infographic literacy among the general audience: while static visualizations are designed to be user-friendly, they still require a certain level of skill to be interpreted effectively. Another factor is that the inherent appeal of static visualizations is insufficient to provoke a response. At the same time, the results show that static visualizations positively mediate the relationship between DJ and conversation. This can suggest that they serve as catalysts for conversation among a specific subset of the audience—those with the expertise and interest to delve deeper into the data.

## 6.3. Limitations and further research

**6.3.1. Single platform.** The study relied on the NYT, as it provided rich content for exploration (Wang & Diakopoulos, 2021). However, focusing on one newspaper and one conversation platform limited the findings to a specific setting and audience, including factors such as manual comment moderation, recommendations and editorial involvement (Masullo et al., 2022). While the uniform moderation practice at NYT ensures consistency in the analyzed data, it may also homogenize certain aspects of the conversation. Further research may examine additional newspapers and conversation platforms, such as X (Twitter), Facebook, Reddit, or Instagram, where different moderation policies and practices may apply. This will enable us to evaluate how the effects differ across platforms, political climates, and cultural contexts.

**6.3.2. Abnormal events.** The study corpus covers a lengthy period, ranging from 2014 to 2022. During this period, there were significant events, such as election campaigns in the US marred by polarization and misinformation, as well as the COVID-19 pandemic, which affected public discourse and social media (Masullo et al., 2022). The impact of such events was not considered in this study. The contribution of data to conversations surrounding these events, and whether it assists in resolving issues, may be examined in further research.

**6.3.3. DJ definition.** The stories were categorized according to the Upshot label provided by the NYT. There is no documented evidence of how the NYT defines DJ, of its relation to DJ literature, or whether it is consistent over time and across all stories. Interpreting the results, we have assumed that the definition is stable. However, NYT definition of DJ may need to be clarified or revised.

**6.3.4. DJ classification.** DJ elements come in a wide variety of types, each corresponding to a different function (Anderson & Borges-Rey, 2019; Stalph, 2018). For example, bar plots and maps are both visualizations, but serve different purposes. Information sources can also be PDF reports or data files. The statistical information can also represent dates. Classifying each element will allow us to understand how they affect the audience and whether additional mediators are involved. Additionally, analyzing the comments' content can provide greater insight into how they are used and perceived (Hullman et al., 2015; McInnis et al., 2020).

**6.3.5. Conversation content.** The degree of conversation is considered to be a measure of its quality (Ksiazek, 2018). Yet, a content analysis is required to determine whether the interactivity addresses data subjects, how it contributes to evidence-based decision-making, and whether it affects other conversations as well (Friess & Eilders, 2015). Analyzing the content allows us to explore the argumentative complexity and whether it supports consensus. In addition, it allows us to examine the characteristics of the target audience and follow their learning process.

## 6.4. Summary

In this study, we examined how growing data accessibility affects audience interactivity. This was explored through DJ and user comments. By analyzing NYT stories from 2014 - 2022 and their comment sections, we discovered that DJ increases the degree of conversation. This relationship is mediated by statistical information, information sources, and static visualizations. In accordance with deliberation theory, this finding provides evidence of the role played by information. Consequently, it is used to engage the audience in conversation, which is a critical stage of democratic processes. Additionally, we demonstrated how secondary news factors contribute to audience engagement. However, fewer comments were made in response to the content, indicating that only a portion of the audience can interpret data effectively. While DJ uses transparency and multimedia to encourage public participation, it appears that only expert audiences are likely to do so. It is unclear whether better DJ design can affect this, or whether it depends on other factors such as audience skills. Several approaches were suggested for further research to address this issue.